\documentclass[reprint,aps, superscriptaddress,nofootinbib]{revtex4-1}

\usepackage{graphicx}
\usepackage{dcolumn}
\usepackage{bm}

\usepackage{amsmath}
\usepackage{amssymb}
\usepackage{amsfonts}
\usepackage{amsthm}
\usepackage{color}
\usepackage{subfig}
\usepackage{MnSymbol}

\begin{document}

\captionsetup{justification=raggedright,singlelinecheck=false}

\title{Anderson localization of composite particles}

\date{\today}

\author{Fumika Suzuki}

 \email{fumika.suzuki@ist.ac.at}
  \affiliation{%
IST Austria (Institute of Science and Technology Austria), Am Campus 1, 3400 Klosterneuburg, Austria
}
  \affiliation{Department of Chemistry, University of British Columbia, Vancouver, British Columbia, Canada V6T 1Z1}
  
  \author{Mikhail Lemeshko}
    \affiliation{%
IST Austria (Institute of Science and Technology Austria), Am Campus 1, 3400 Klosterneuburg, Austria
}

  \author{Wojciech H. Zurek}
    \affiliation{%
Theory Division, Los Alamos National Laboratory, Los Alamos, New Mexico 87545, USA
}

\author{Roman V. Krems}
  \affiliation{Department of Chemistry, University of British Columbia, Vancouver, British Columbia, Canada V6T 1Z1}
  \affiliation{Stewart Blusson Quantum Matter Institute, University of British Columbia, Vancouver, British Columbia, Canada V6T 1Z4}
    
  \begin{abstract}
We investigate the effect of  coupling between translational and internal degrees of freedom of composite quantum particles on their localization in a random potential. We show that entanglement between the two degrees of freedom weakens localization due to the upper bound imposed on the inverse participation ratio by purity of a quantum state. We perform numerical calculations for a two-particle system bound by a harmonic force in a 1D disordered lattice and a rigid rotor in a 2D disordered lattice. We illustrate that the coupling has a dramatic effect on localization properties, even with a small number of internal states participating in quantum dynamics. 
\end{abstract}

\maketitle


Anderson localization (AL) \cite{anderson}, or lack thereof, determines propagation of waves through disordered media. 
As such, it has  been explored in a wide range of contexts, 
including quantum thermalization \cite{manybody, ser}, quantum walks \cite{roman, marek}, complex networks and graphs \cite{graph, graph2}, open system dynamics \cite{bath,open, open2, open3, transport}, quantum chaos \cite{chaos, suzuki}, and adiabatic quantum computation \cite{alt}. While Anderson originally studied non-interacting particles in disordered crystalline lattices, recent work has elucidated the effect of interactions \cite{interactions, interactions2, interactions3, interactions4, interactions5, interactions6, interactions7, interactions8},  dimensionality \cite{dimensionality, dimensionality2}, and hopping range \cite{roman2}, demonstrating interesting phenomena such as cooperative shielding \cite{cooperative-shielding}, making particles with long-range hopping localize effectively as those with short-range hopping. The fundamental importance of  localization of quantum particles has stimulated the development of a wide range of experimental platforms aiming to observe AL directly \cite{experiment3,  experiment2}, culminating in the imaging of Anderson-localized states with ultracold atoms in optical lattices \cite{ultracold-atoms, ultracold-atoms2}.  At the same time, AL was re-examined in the context of coherent energy transfer in photosynthetic light-harvesting systems, organic photovoltaics, conducting polymers and $J$-aggregate thin films \cite{moix, moix2, moix3, moix4, moix5, moix6, moix7, moix8, moix9,moix10}. In such systems, energy is carried by excitons and exciton-polarons, which may undergo localization.

While previous studies considered AL of structureless particles, these recent experiments suggest the possibility of observing AL of quantum particles with internal structure. 
It has now become possible to trap ultracold molecules in optical lattices \cite{molecules-in-lattice}. This paves the way for studying the effects of molecular ro-vibrational structure on AL of ultracold molecules. It was also demonstrated that excitons may form bound pairs, even in the Frenkel exciton limit \cite{biexciton, suzuki2}, where exciton pairs are comparable in size to lattice spacing. 
Similarly, polarons in conducting polymers may bind into bipolarons with light effective mass \cite{bipolaron-references, john-2}. The quantum behavior of those composite  particles can be strongly affected by their internal degrees of freedom \cite{suzuki2, suzuki4, suzuki3,unruh, flam, arndt, arndt2}. Although quantum transport of structureless particles coupled to an external environmental bath has been well studied \cite{moix},
 the effect of internal dynamics of such bound pairs on AL has not been thoroughly investigated. Understanding it is of key importance for the prospects of organic photovoltaics and bipolaronic superconductivity in organic materials.

Here, we study the effect of coupling between translational motion and internal states of composite quantum particles on their localization. We first present general arguments illustrating delocalization induced by coupling to internal states. 
We formulate the problem as localization of states in space of one of subsystems of a composite system. This allows us to derive the limits imposed on the localization by  coupling with the other subsystem.
We then perform numerical calculations for two model systems illustrated in Fig.  1: a two-particle system bound by a harmonic force in a one-dimensional (1D) disordered lattice and a rigid rotor of two particles in a two-dimensional (2D) disordered lattice. 

\begin{figure}
{%
\includegraphics[clip,width=0.4\columnwidth]{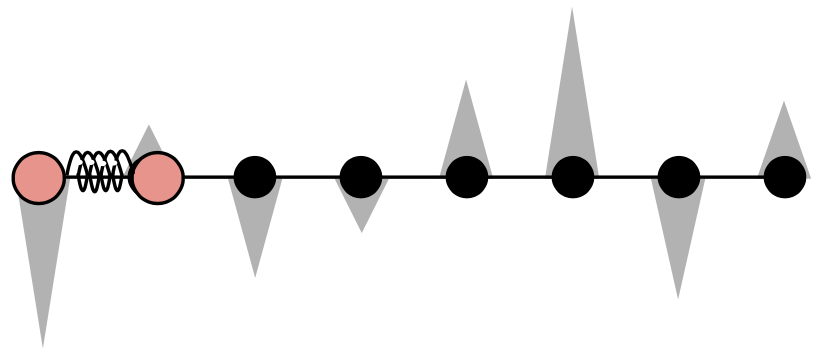} 
\includegraphics[clip,width=0.35\columnwidth]{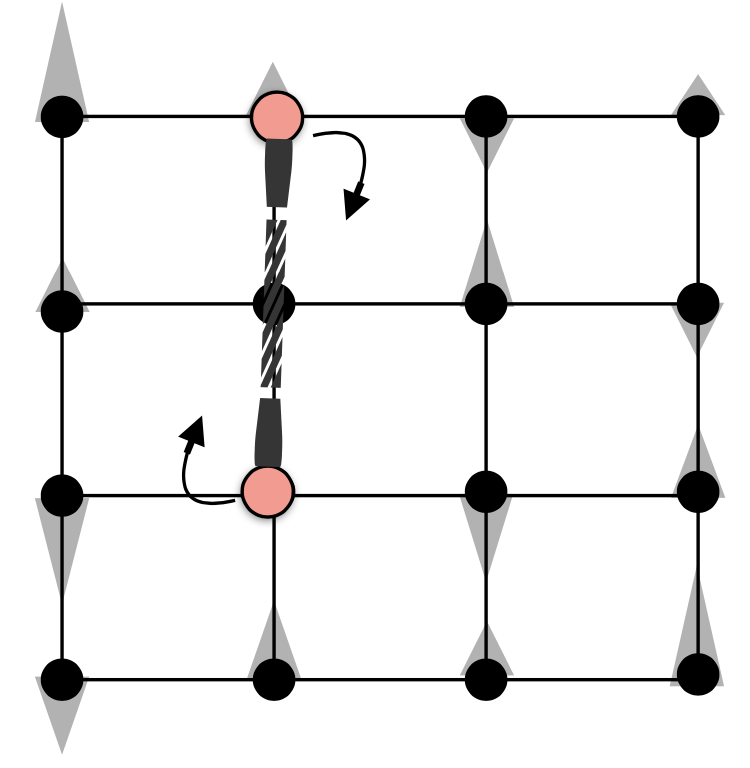}%
}
\caption{Models used for numerical calculations: left -- harmonically interacting  particles in a 1D disordered lattice; 
right -- rigid rotor of two particles separated by two lattice constants in a 2D disordered lattice. 
}
\label{fig1}
\end{figure}

The problem considered here is different from both the conventional Anderson model and from many-body localization (MBL)  \cite{manybody, ser}. 
As in the conventional Anderson model, we consider a single non-interacting particle in a disordered lattice. The dynamics of the composite particle is, however, affected by couplings to internal states that provide which-way information, that may suppress quantum interferences thus affecting localization in  the spatial dimension. In this sense, the present problem is related to Anderson localization in a quantum system coupled to an environment. However, in contrast to problems with baths, the dimension of the internal state space is small and the two coupled subsystems are mutually determined by the dynamics of the composite particle as a whole. If viewed as a wave packet in the combined space of spatial and internal degrees of freedom,  the  present problem represents quantum dynamics on a Cartesian product of a lattice graph and a complete graph associated with  spatial and internal degrees of freedom respectively. This could be contrasted to MBL, whose Fock-space graphs are high-dimensional and maximally-correlated \cite{graphs}. 


We  represent 
a  pure state of a composite system by a wavefunction $\Psi_{SE}(\mathbf{R},\mathbf{n}) \in \mathcal{H}_{S}(\mathbf{R})\otimes\mathcal{H}_{E}(\mathbf{n})$ with the Hilbert spaces $\mathcal{H}_{S}$ and $\mathcal{H}_{E}$ of dimensionality $d_{S}=\mbox{dim}(\mathcal{H}_S)$ and $d_{E}=\mbox{dim}(\mathcal{H}_{E})$, respectively. For a composite quantum particle considered in this article, the subsystems $S$ and $E$ describe the translational position by $\bf R$ and the internal degrees of freedom by $\bf n$, respectively.
The reduced density matrix of $S$ in coordinate space $\bf R$ after tracing out $E$ is written as
\begin{eqnarray}
\rho_{S}(\mathbf{R},\mathbf{R'})=\displaystyle \sumint_{\mathbf{n}}\rho_{SE}(\mathbf{R},\mathbf{R}';\mathbf{n},\mathbf{n})
\end{eqnarray}
where the sum $\sum$ is used for discrete states and the integral -- for continuum. Localization of the subsystem $S$ in coordinate space $\mathbf{R}$ is quantified by the inverse participation ratio (IPR)
defined as
\begin{eqnarray}
\xi =\displaystyle\sum_{\mathbf{R}}|\rho_{S}(\mathbf{R},\mathbf{R})|^2
\end{eqnarray}
where $\displaystyle\lim_{d_{S}\rightarrow\infty}\xi\rightarrow 1/d_{S}$ for extended states, while $\xi$ is constant $\gg1/d_{S}$ for localized states.  When only one lattice site is occupied, we have 
$\xi=1$. 

The mixing of $S$ and $E$ is quantified by purity defined as $\gamma  = \mbox{tr}\rho_{S}^2$. It can be seen that the purity $\gamma$ is related to  $\xi$ by the following expression:
\begin{eqnarray}\label{gamma}
\gamma =\mbox{tr}\rho_{S}^2=\xi +\displaystyle\sum_{\mathbf{R}\not=\mathbf{R}'}|\rho_{S}(\mathbf{R},\mathbf{R}')|^2 \geq \xi.
\label{gamma-xi}
\end{eqnarray}

The value of $\gamma$ thus puts an upper limit on $\xi$, which can be used to elucidate the effect of mixing between $S$ and $E$ on localization in $S$. Since $1/d \leq \gamma \leq 1$, where $d= \mbox{min}(d_{S},d_{E})$, we have $\xi\leq 1$ for a pure state, while $\xi \leq 1/d$ for a completely mixed state. 
When $S$ is strongly entangled with a large number of degrees of freedom, with $d_{E}\geq d_{S}$, we have fully extended states with $ \xi = 1/d_{S}$.

This limit is particularly relevant for a quantum particle $S$ coupled to a  bath $E$ with a large number of degrees of freedom. In this case, $d_{E}\geq d_{S}$ and the decoherence \cite{dec, zurek} induces extended states when $S$ is maximally entangled with the bath. This result is consistent with the interpretation of AL as a consequence of interference between multiple scattering paths \cite{path}. If the scattering particle is coupled to a bath causing decoherence, this interference is destroyed and the scattering particle must effectively diffuse as a classical particle. Eq. (\ref{gamma-xi}) quantifies this argument by showing that localization would be destroyed when the number of accessible degrees of freedom in the bath is  larger than $d_S$. 


\begin{figure*}
{%
\includegraphics[clip,width=2\columnwidth]{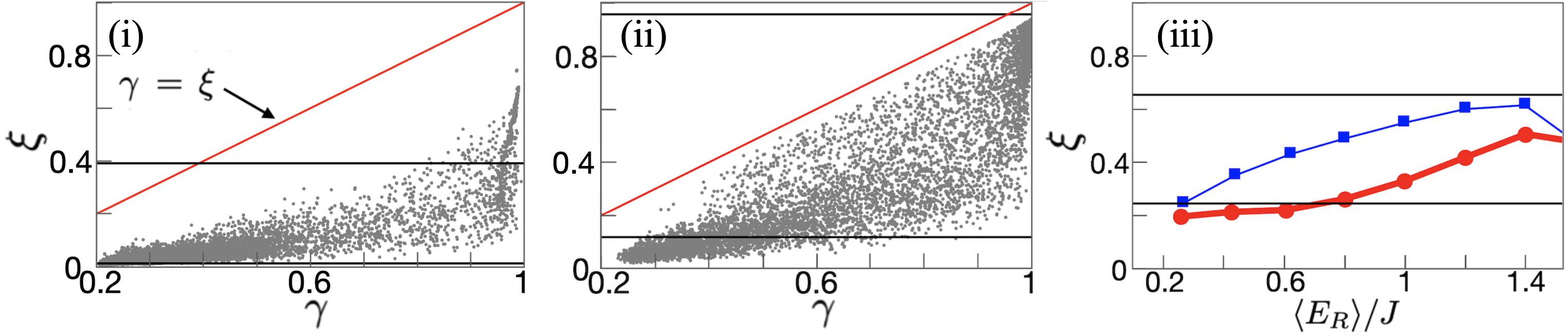}%
}
\caption{
(i, ii) $\xi$ and $\gamma$ of  eigenstates of  (\ref{ham2}); (iii) IPR averaged within energy bins for a composite particle (red line, circles) and for a structureless particle (blue line, squares). The calculation parameters are $\omega=0.1|J|$, 5 internal states ($n=0,2,4,6,8$),  $\lambda=|J|$ for (i) and $\lambda=5|J|$ for (ii, iii).
The horizontal  solid black lines show (i, ii) the minimum and maximum values of $\xi$  and (iii) the averaged $\xi$ for a structureless particle from the entire spectrum. All plots are produced with $N_R = 200$ sites and 20 disorder realizations. 
}
\label{fig2}
\end{figure*}

\begin{figure}
{%
\includegraphics[clip,width=0.75\columnwidth]{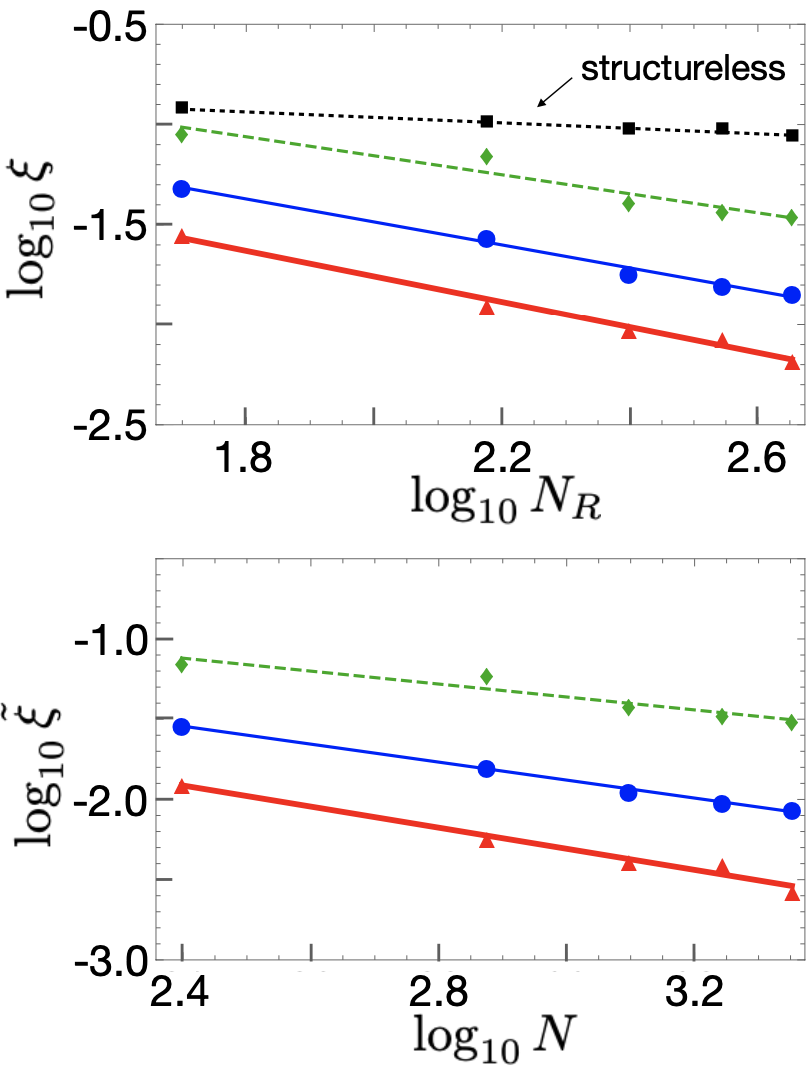}%
}
\caption{
$\log_{10} \xi$ (upper panel) and $\log_{10}\tilde{\xi}$ (lower panel) of the most extended eigenstate  of  (4) with $\langle E_{R}\rangle <1.5|J|$, $\lambda=3|J|$, 5 internal states ($n=0,2,4,6,8$), $\omega=0.1|J|$ (average purity $\bar{\gamma}=0.24$) for triangles (red), $\omega=0.6|J|$ ($\bar{\gamma}=0.48$) for circles (blue) and $\omega=1.2|J|$ ($\bar{\gamma}=0.84$) for diamonds (dashed, green), as functions of $\log_{10} N_{R}$, where the number of lattice sites $N_{R}\in [50,450]$ (upper panel) and $\log_{10} N$ where $N\in [50\times 5, 450 \times 5]$  (lower panel). The results are averaged  over 20 disorder realizations. The dotted black line in the upper panel represents the IPR scaling of the most extended state from the entire spectrum of a structureless particle in the same 1D disordered lattice.
}
\label{fig3}
\end{figure}


More generally, Eq. (\ref{gamma-xi}) shows that the localization in $S$ can be weakened even when the environment space has a lower dimension than that of $S$. We define the ratio $\Delta=\xi/\gamma$ to characterize quantum states. The value 
$\Delta \sim 1/d_S$ is characteristic of states delocalized already in the absence of $E$, 
while states with up to $\Delta \sim d/d_S$ can exhibit delocalization induced by the entanglement, i.e., there can exist both localized states with $\gamma\sim1$ and delocalized states with $\gamma <1$ in this regime. On the other hand, states with $d/d_S 
\ll \Delta \leq 1$ exhibit  localization even with the coupling to $E$, while their IPR can be reduced due to the upper limit given by $\gamma$. In what follows, we numerically demonstrate that coupling of the translational motion with internal degrees of freedom of composite particles  can weaken their localization and can induce the delocalization of the type $\Delta \sim d/d_S$ for a rigid rotor in a 2D disordered lattice.

\begin{figure*}
{%
\includegraphics[clip,width=1.9\columnwidth]{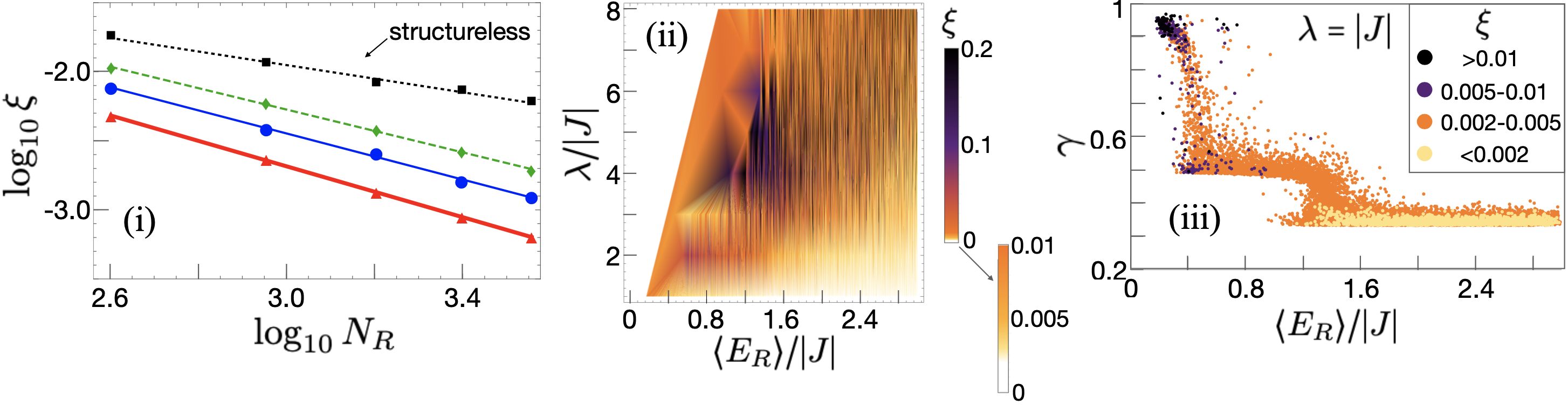}%
}
\caption{ 
  (i) $\log_{10} \xi$  of the most extended eigenstate  of  (\ref{hamrotor2}) with $\langle E_{R}\rangle <3|J|$, $\lambda=4|J|$, $1/r^2=0.1|J|$ (average purity $\bar{\gamma}=0.36$) for triangles (red), $1/r^2=0.25|J|$ ($\bar{\gamma}=0.42$) for circles (blue) and $1/r^2=|J|$ ($\bar{\gamma}=0.94$) for diamonds (dashed, green), as functions of the logarithm of the number of lattice sites $N_R \in [20^2,60^2]$. The dotted black line represents the IPR scaling of the most extended state from the entire spectrum of a structureless particle in the same 2D disordered lattice.
(ii) $\xi$ of eigenstates of  (\ref{hamrotor2})  as a function of  $\langle E_R\rangle$ and disorder strength $\lambda$. (iii) $\xi$, $\gamma$ and $\langle E_R\rangle$ of the eigenstates when $\lambda=|J|$. All plots are obtained by averaging over 20 disorder realizations, and for 3 internal states ($n=0,2,4$). $N_R=30^2$ for (ii) and (iii).}.
\label{fig4}
\end{figure*}

 We now present numerical calculations. First, we study a quasi-one-dimensional Anderson model for two particles on a disordered lattice bound by a harmonic force ({\it cf.}, Fig. \ref{fig1}, left). 
In this case, ${\bf n}=n$ represents the vibrational states in the two-particle dynamics. 
Following \cite{unruh, suzuki3}, the relevant Hamiltonian is
\begin{eqnarray}\label{ham2}
&&H=J\displaystyle\sum_{R} (\hat{c}^{\dagger}_{R+1,n}\hat{c}_{R,n}+\hat{c}^{\dagger}_{R-1,n}\hat{c}_{R,n})\nonumber\\
&&+\displaystyle\sum_{R,n}(-2J +E_{n})\hat{c}^{\dagger}_{R,n}\hat{c}_{R,n}+\displaystyle\sum_{m,n,R}V_{nm}(R)\hat{c}^{\dagger}_{R,n}\hat{c}_{R,m}
\end{eqnarray}
where $\hat{c}_{R,n}^{\dagger}$ is the creation operator of a two-particle system (i.e., composite particle) with the lattice-site position $R$ and the internal vibrational state $n$, $J$ is the hopping  amplitude and $E_{n}=2\omega (n+1/2)$ is the energy of the internal states with the frequency $\omega$. $V_{nm} (R)$ is the effective potentials determined by the external random potential and the wavefunction of the internal states, $V_{nm}(R)=\displaystyle\sum_{l\in\mathbb{Z}}\lambda_{l}(\phi_{n}(2l-R)\phi_{m}(2l-R)+\phi_{n}(R-2l)\phi_{m}(R-2l))$ where $\phi_{n}(r) =((\omega /\pi)^{1/4}/\sqrt{2^n n!}) H_{n} (\sqrt{\omega}r)\exp (-\omega r^2/2)$, and $\lambda_{l}$ are random variables from a uniform distribution over $\left [ -\lambda, \lambda \right ]$. We see that $V_{nm}(R)=0$ unless $n,m$ are both even or both odd. Thus, interactions with disorder lead to transitions between the internal states of the same parity. The Hamiltonian is derived in the low energy regime (see supplemental material for the derivation). In the following, we present numerical results for a composite particle with $\langle E_{R} \rangle <1.5|J|$ and low energy  internal oscillator states.

Fig.  \ref{fig2} (i, ii) shows $\xi$ and $\gamma$ of  eigenstates  of (\ref{ham2}) and Fig.  \ref{fig2} (iii) -- IPR averaged within energy bins for multiple disorder realizations  with (i) $\lambda =|J|$ and (ii, iii) $\lambda =5|J|$. We set $N_R=200$ sites, $\omega=0.1|J|$ and allow 5 vibrational states with even $n=0,2,4,6,8$. This gives the minimum purity $1/5$. The  expectation value of the translational energy is given by $\langle E_{R}\rangle=\langle H_{R}\rangle$ where $H_{R}$ is given by the first three terms of Eq. (\ref{ham2}). For a single structureless particle, the eigenenergy of the Anderson model is directly related to its group velocity which affects its localization properties. However, the eigenenergy of (\ref{ham2})  additionally contains contributions from the internal states of the composite particle. For this reason, we use $\langle E_{R}\rangle $ which can be related to the group velocity of the translational motion of the composite particle, so that the results can be directly compared to those of the structureless particle with a similar group velocity.

In Fig.  \ref{fig2}  (i, ii), as the disorder strength $\lambda$  increases, the distribution approaches -- but never exceeds -- the upper bound $\xi = \gamma$ represented by the  solid red line.   
This shows that, for the composite particle with $\gamma <1$, there exists a limitation on the localization strength even when the disorder is very strong, i.e., $\lambda \gg |J|$. This can be further confirmed by comparing the maximum IPR for a structureless particle and that for the composite particle with $\gamma <1$ in a strong disorder.  Fig.  \ref{fig2}  (iii) shows that  the averaged IPR for a composite particle  is lower than that for a structureless particle. The horizontal solid lines represent (i, ii) the maximum and minimum of the IPR of all eigenstates and (iii) the averaged IPR,  from the entire spectrum of a structureless  particle in a 1D lattice with the same disorder. The eigenstates for a composite particle with small $\gamma$ have suppressed localization compared to a structureless particle. However some eigenstates with $\gamma \sim 1$  can exhibit stronger localization than those of a structureless  particle. This happens because, when two parts of a composite particle are in proximity, the effective random potential can become larger than that for a structureless particle.

 Eq. (\ref{ham2}) effectively describes a (1+$\epsilon$)-dimensional system with $\epsilon$ accounting for internal states. The coupling to internal states is controlled by the value of $\omega$. Generally, as $\omega$ increases, transitions between different internal states become less likely to occur, and  the problem reduces to a 1D problem for each internal state. This can be seen in the increase of the purity of the eigenstates as an indication of the separability of the Hamiltonian in  translational and internal degrees of freedom. Fig. \ref{fig3} shows the scaling of $\xi$ and $\tilde{\xi}$ of the most extended state with lattice size for different values of $\omega$. Here $\tilde{\xi}$ is the IPR computed without tracing out the internal degrees of freedom, i.e., $\tilde{\xi}=\displaystyle\sum_{\mathbf{R},\mathbf{n}}|\rho_{SE} (\mathbf{R},\mathbf{R};\mathbf{n},\mathbf{n})|^2$. We observe that $\tilde{\xi}$ approaches $\xi$ as $\omega$ and  purity  increase, indicating that the Hamiltonian becomes almost separable in translational and internal spaces, and the $(1+\epsilon)$-dimensional problem is nearly reduced to a 1D problem.  For low $\omega$ (and small purity), the IPR of the oscillator is fairly small compared to that of a structureless particle. However,  the extended states are not observed for the 1D oscillator.

Second, we consider a quasi-two-dimensional Anderson model which describes  dimers undergoing rigid rotor dynamics on a two-dimensional lattice, as shown in Fig.  1. 
The relevant Hamiltonian is
\begin{eqnarray}\label{hamrotor2}
&&H=J\displaystyle\sum_{x,y,n}(\hat{c}^{\dagger}_{x+1,y,n}\hat{c}_{x,y,n}+\hat{c}^{\dagger}_{x,y+1,n}\hat{c}_{x,y,n}+h.c.)\nonumber\\
&&+\displaystyle\sum_{x,y,n}(-4J+E_{n})\hat{c}^{\dagger}_{x,y,n}\hat{c}_{x,y,n}+\displaystyle\sum_{m,n,x,y}V_{nm}(x,y)\hat{c}^{\dagger}_{x,y,n}\hat{c}_{x,y,m}\nonumber\\
\end{eqnarray}
where $\hat{c}^{\dagger}_{x,y,n}$ is the creation operator of the rigid rotor with the translational position $\mathbf{R}=(x,y)$ and the internal rotational state $n$, $J$ is the lattice hopping amplitude and $E_{n}=n^2/r^2$ is the energy of the internal states with a fixed distance $r$ between two particles of the rotor. $V_{nm}(R)$ is the effective potential determined by the disorder potential and the wavefunction of the internal states, $V_{nm}(x,y)=\displaystyle\sum_{l,l'\in\mathbb{Z}}\lambda_{l,l'}(\phi_{n}^{*}(\theta_{xy})\phi_{m}(\theta_{xy})+\phi_{n}^{*}(\theta'_{xy})\phi_{m}(\theta'_{xy}))$
with $\theta_{xy}=\mbox{arctan}\left(\frac{2l'-y}{2l-x}\right)$,  $\theta'_{xy}=\mbox{arctan}\left(\frac{2l'-y}{2l-x}\right)-\pi$ and $|2l'-y|\leq r$, $|2l-x|\leq r$. The quantum states of the rotor are given by $\phi_{n}(\theta)=e^{in\theta}/\sqrt{2\pi}$, while $\lambda_{l,l'}$ are random variables with a uniform distribution from $-\lambda$ to $\lambda$. $V_{nm}(x,y)=0$ unless $(n,m)$ are both even or both odd. We show the numerical results for a composite particle with $ \langle E_R\rangle < 3|J|$ and low energy  rotational states, where the present Hamiltonian is valid (see SM for the derivation details).



As in the previous example, Fig.  \ref{fig4} (i) shows that the scaling is markedly different for structureless particles and the 2D rotor when $\gamma$ is small. It is demonstrated that, 
for the 2D rotor, a coupling to just three rotational states accelerates the scaling of $\xi$ to a great extent, indicating delocalized states of the type $\Delta \sim d/d_S$. Fig.  \ref{fig4} (ii) shows $\xi$ of eigenstates of (\ref{hamrotor2}) as a function of $\langle E_R\rangle$ and disorder strength $\lambda$. It  can be seen that the states tend to become delocalized as $\lambda$ decreases and $\langle E_R\rangle$ increases. Note that the increase of $\xi$ for large $\lambda$ slows down as the distribution of the eigenstates approaches the upper bound $\xi=\gamma$. However, purity is also important in the localization properties of a composite particle, as seen in Fig.  \ref{fig4} (iii), illustrating that even with similar $\langle E_R\rangle$, the eigenstates with smaller $\gamma$ tend to be more delocalized. This demonstrates that the rotor in a 2D disordered lattice exhibits rich complex behavior compared to that of a structureless particle in a 3D disordered lattice whose localization properties are mainly determined by its energy, which leads to the simple separation of localized  and extended states by the mobility edge.

 In summary, we have shown that the coupling between the translational and internal degrees of freedom weakens the localization of composite particles in disordered lattices. The internal degrees of freedom can be viewed as a small quantum system coupled to the translational degrees of freedom, which suppresses localization as an interference phenomenon. We have shown that the upper bound of $\xi$ given by $\gamma$ imposes the limitation on the localization strength even at strong disorder. The internal degrees of freedom can reduce localization  or induce extended states. In both cases, the effect of the internal degrees of freedom becomes remarkable when purity is small. This happens when  the translational energy is comparable to the characteristic energy of internal states so that effective energy transfer between two subspaces is significant and the Hamiltonian becomes inseparable in the two degrees of freedom.
To support our conclusions, we have presented numerical results for quantum particles with vibrational motion in 1D disordered lattices and rigid rotor dynamics in 2D disordered lattices. Our results illustrate that coupling to just three rotational states of a rigid rotor on a 2D lattice changes dramatically the lattice-size scaling properties of translational states, inducing the formation of extended states. 
\\

\begin{acknowledgements}

We acknowledge helpful discussions with W. G. Unruh and A. Rodriguez. F.S. is supported by  the European Union’s Horizon 2020 research and innovation programme under the Marie Skłodowska-Curie Grant No. 754411. M.L. acknowledges support by the European Research Council (ERC) Starting Grant No. 801770 (ANGULON). W.H.Z is supported by Department of Energy under the Los Alamos National Laboratory LDRD Program as well as by the U.S. Department of Energy, Office of Science, Basic Energy Sciences, Materials Sciences and Engineering Division, Condensed Matter Theory Program. R.V.K. is supported by NSERC of Canada.

\end{acknowledgements}

\section*{Supplemental material}

\subsection*{Derivation of the Anderson model for a harmonic oscillator in 1D}

\setcounter{equation}{0}
\renewcommand{\theequation}{A\arabic{equation}}

We start with the Hamiltonian in the continuum space which describes two particles with unit mass bound by a harmonic potential in a one-dimensional space,
\begin{eqnarray}\label{A1}
H_0=\frac{p_1^2}{2}+\frac{p_2^2}{2}+ \omega^2 (x_1-x_2)^2.
\end{eqnarray}

Using the translational position $R=x_1 +x_2$, and the relative distance $r=x_1-x_2$, the eigenstates of the Hamiltonian can be written as $\Phi (R) \phi_{n}(r)$ where
\begin{eqnarray}
\phi_{n}(r) &=&((\omega /\pi)^{1/4}/\sqrt{2^n n!}) H_{n} (\sqrt{\omega}r)\exp (-\omega r^2/2),\nonumber\\
E_{n}&=&2\omega (n+1/2).
\end{eqnarray}

We introduce the random potential as follows
\begin{eqnarray}
V(x_1,x_2)=\displaystyle\sum_{l\in \mathbb{Z}}\lambda_{l} (\delta (x_1-l)+\delta(x_2 -l))
\end{eqnarray}
where $\lambda_{l}$ is a random variable and $\delta (x-l)$ is the delta function. In terms of $R$ and $r$, the random potential can be written as
\begin{eqnarray}
V(R,r)=\displaystyle\sum_{l\in \mathbb{Z}}\lambda_{l}(\delta (R+r-2l)+\delta (R-r-2l)).
\end{eqnarray}

We expand the eigenstates of $H=H_0+V$ as
\begin{eqnarray}
\Psi =\displaystyle\sum_{n} u_{n} (R)\phi_{n}(r).
\end{eqnarray}

This reduces the time-independent Schr\"{o}dinger equation to
\begin{eqnarray}\label{eq}
-\frac{d^2 u_{n}(R)}{d^2 R}+E_{n}u_{n}(R)+\displaystyle\sum_{m}V_{nm}(R)u_{m}(R)=Eu_{n}(R)\nonumber\\
\end{eqnarray}
where
\begin{eqnarray}
V_{nm}(R)&=&\displaystyle\sum_{l\in\mathbb{Z}}\lambda_{l}(\phi_{n}(2l-R)\phi_{m}(2l-R)\nonumber\\
&&\qquad\quad+\phi_{n}(R-2l)\phi_{m}(R-2l)).
\end{eqnarray}

By discretizing (\ref{eq}) using finite-difference methods, we write the Hamiltonian
\begin{eqnarray}\label{ham}
H&=&J\sum_{R}(|R+1\rangle \langle R|+|R-1\rangle\langle R|)\otimes I_{n}\nonumber\\
&&+(-2J +E_{n})I_{R}\otimes I_{n}+\sum_{m,n,R}V_{nm}(R)|R,n\rangle\langle R,m|\nonumber\\
\end{eqnarray}
acting on the state
\begin{eqnarray}
|\Psi\rangle =\sum_{R,n} u_{n} (R)|R, n\rangle
\end{eqnarray}
where $J=-1/a^2$ with the lattice spacing $a$, and we define
\begin{eqnarray}
|R,n\rangle =\displaystyle\sum_{r} \phi_{n}(r)|R,r\rangle=\hat{c}^{\dagger}_{R,n}|0\rangle.
\end{eqnarray}

Here, $\hat{c}_{R,n}^{\dagger}$ is the creation operator of a composite particle with the translational position $R$ and the internal vibrational state $n$.  The Hamiltonian (\ref{ham}) can then be written as
\begin{eqnarray}\label{HamA}
H&=&J\displaystyle\sum_{R} (\hat{c}^{\dagger}_{R+1,n}\hat{c}_{R,n}+\hat{c}^{\dagger}_{R-1,n}\hat{c}_{R,n})\nonumber\\
&&+\displaystyle\sum_{R,n}(-2J +E_{n})\hat{c}^{\dagger}_{R,n}\hat{c}_{R,n}+\displaystyle\sum_{m,n,R}V_{nm}(R)\hat{c}^{\dagger}_{R,n}\hat{c}_{R,m}.\nonumber\\
\end{eqnarray}

An alternative approach to arrive at (\ref{HamA}) is to start with the Hamiltonian which describes two particles interacting with each other in a  disordered lattice,
\begin{eqnarray}\label{hamsub}
H&=&\displaystyle\sum_{i} (-4J' \hat{a}^{\dagger}_{i}\hat{a}_{i}+J' (\hat{a}_{i+1}^{\dagger}\hat{a}_{i}+\hat{a}^{\dagger}_{i-1}\hat{a}_{i}))\nonumber\\
&&+\displaystyle\sum_{i,j}U (|i-j|) a_{i}^{\dagger}a_{j}^{\dagger}a_{j}a_{i}+\displaystyle\sum_{i}V_{i}\hat{a}_{i}^{\dagger}\hat{a}_{i}
\end{eqnarray}
where $\hat{a}_{i}$ is the creation operator for a particle in site $i$, $U(|i-j|)$ is the interaction strength between two particles and $V_{i}$ is the on-site  potential with a random distribution.

In principle, one can diagonalize (\ref{hamsub}) and select the eigenstates which are bounded for the relative distance between two particles to investigate the localization of composite quantum particles. However, we can reduce dimensionality  by performing the following procedure to simplify the numerical computations. 

By introducing the translational position $R=i+j$, and the relative distance $r=i-j$, we  rewrite (\ref{hamsub}) as
\begin{eqnarray}\label{hamsub2}
H&=&-4J'\displaystyle\sum_{R,r}|R,r\rangle\langle R,r| +J'(\hat{\frak{R}}+\hat{\frak{R}}^{\dagger})(\hat{\frak{r}}+\hat{\frak{r}}^{\dagger})\nonumber\\
&&+\displaystyle\sum_{R,r}U(|r|)|R,r\rangle\langle R,r|\nonumber\\
&&+\displaystyle\sum_{l}V_{l}\displaystyle\sum_{R,r} (\delta (R+r-2l)+\delta (R-r-2l))|R,r\rangle\langle R,r|\nonumber\\
\end{eqnarray}
where $\hat{\mathfrak{R}}=\displaystyle\sum_{R,r}|R+1,r\rangle\langle R,r|$, $\hat{\mathfrak{r}}=\displaystyle\sum_{R,r}|R,r+1\rangle\langle R,r|$,  and $U(|r|)=\omega^2 r^2$ for a harmonic potential.

The first two terms of (\ref{hamsub2}) give the kinetic energy for a composite particle, $E_0=-4J'+4J'\cos aK \cos ak$ where $K$ and $k$ are the wave vectors associated with the translational motion and the relative motion respectively. In the low energy limit, we  can introduce the Taylor expansion to obtain 
\begin{eqnarray}
E_0\approx -2J' (a^2 k^2+ a^2 K^2).
\end{eqnarray}
The parabolic dispersion mimics the cosine dispersion with $Ka, ka < \pi/2$. Therefore, the wavefunction for  harmonic oscillators on a lattice can be approximately obtained by the Hamiltonian  in  the continuum space (\ref{A1}),  and the problem reduces to the one described by (\ref{HamA}) with $J'=J/2$ when we project the Hamiltonian onto the  set of states of the harmonic oscillators $|R,n\rangle$.

This indicates that (\ref{HamA}) gives a good approximation when there exist lattice points within an associated length scale (i.e., wavelength)   so that the spatial form of the wavefunction can be realized  on a lattice. For the harmonic oscillator, the length scale is typically $4A_{n}/(n+1)$ where the amplitude of the oscillation, $A_{n}=\sqrt{(2n+1)/\omega}$. 

In this article, we investigate the numerical results in the regime $Ka< \pi/2$. Furthermore, the condition $4A_{n}/(n+1) > a$ is satisfied for all $n$ considered in this work.

\subsection*{Derivation of the Anderson model for a rigid rotor in 2D}

\setcounter{equation}{0}
\renewcommand{\theequation}{B\arabic{equation}}
\setcounter{figure}{0}  
\renewcommand\thefigure{B\arabic{figure}}    

\begin{figure*}
{%
\includegraphics[clip,width=1.9\columnwidth]{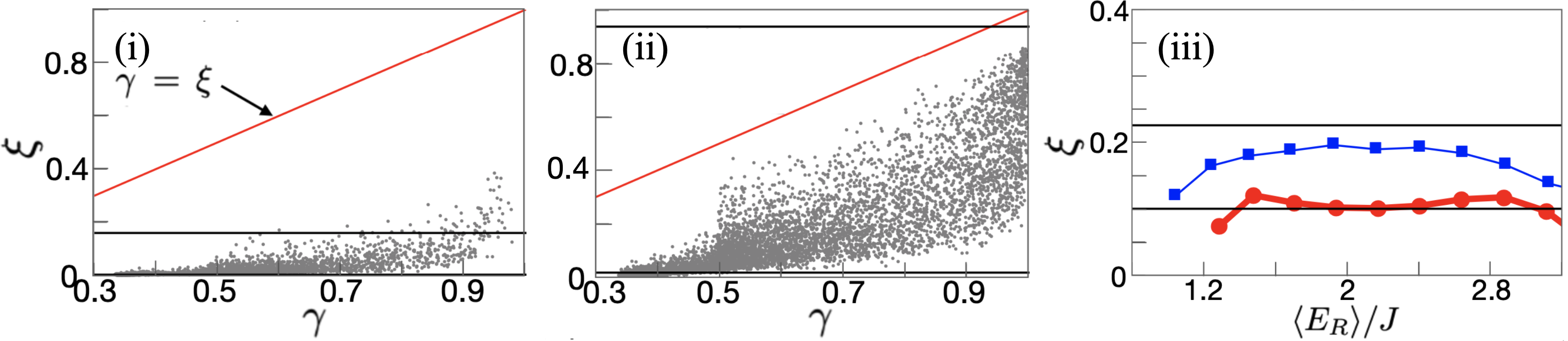} 
}
\caption{(i, ii) $\xi$ and $\gamma$ of  eigenstates of  (\ref{hamsupp}); (iii) IPR averaged within energy bins for a composite particle (red line, circles) and for a structureless particle (blue line, squares). The calculations parameters are $1/r^2=0.1|J|$, 3 internal states ($n=0,2,4$),  $\lambda=2|J|$ for (i) and $\lambda=5|J|$ for (ii, iii).
The horizontal  solid black lines show (i, ii) the minimum and maximum values of $\xi$  and (iii) the averaged $\xi$ for a structureless particle from the entire spectrum. All plots are produced with $N_R = 30^2$ sites and 20 disorder realizations.
}
\label{fig5}
\end{figure*}

\begin{figure}
{%
\includegraphics[clip,width=0.9\columnwidth]{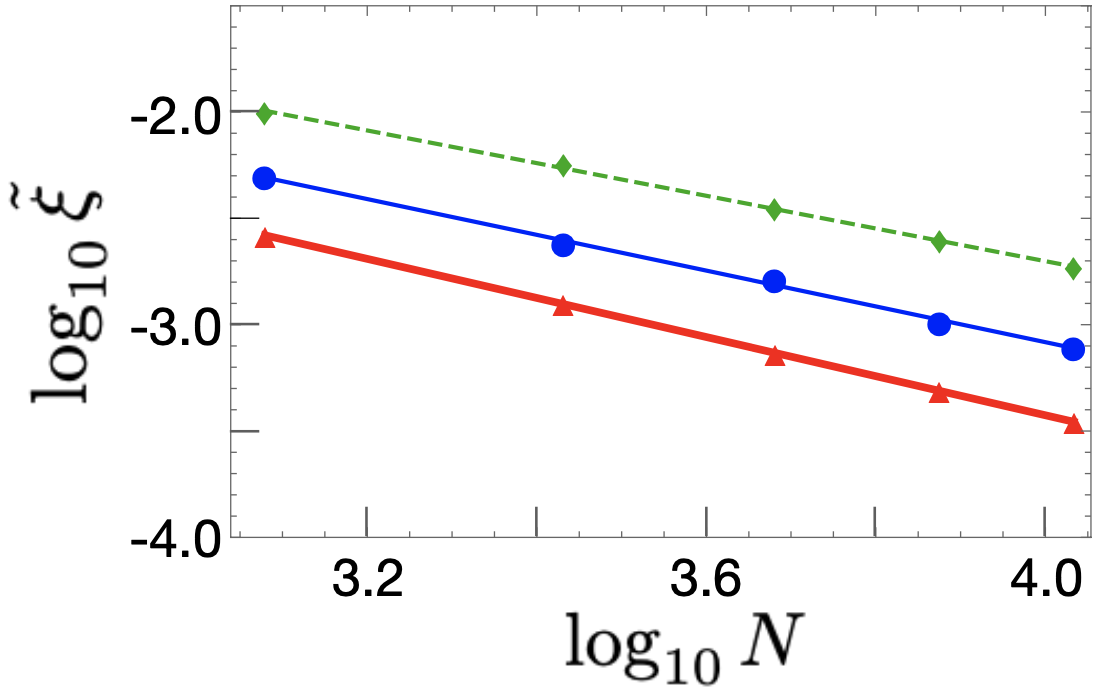} 
}
\caption{ $\log_{10} \tilde{\xi}$  of the most extended eigenstate  of  (\ref{hamsupp}) with $\langle E_{R}\rangle <3|J|$, $\lambda=4|J|$, 3 internal states ($n=0,2,4$), $1/r^2=0.1|J|$ (average purity $\bar{\gamma}=0.36$) for triangles (red), $1/r^2=0.25|J|$ ($\bar{\gamma}=0.42$) for circles (blue) and $1/r^2=|J|$ ($\bar{\gamma}=0.94$) for diamonds (dashed, geen), as functions of  $\log_{10} N$, where  $N \in [20^2\times 3,60^2\times 3]$. The results are averaged  over 20 disorder realizations. 
}
\label{fig6}
\end{figure}

For a rigid rotor, we start with the Hamiltonian which describes two rigidly bound particles in a two-dimensional space:
\begin{eqnarray}\label{rotor0}
H_0=\frac{\mathbf{p}_1^2}{2}+\frac{\mathbf{p}_2^2}{2}+U(|\mathbf{r}_1-\mathbf{r}_2|)
\end{eqnarray}
where $\mathbf{r}_1=(x_1,y_1)$ and $\mathbf{r}_2 =(x_2,y_2)$ represent the positions of two particles respectively.

As before, we introduce the translational coordinate $\mathbf{R}=\mathbf{r}_1+\mathbf{r}_2=(x,y)$ and the relative distance $\mathbf{r}=\mathbf{r}_1-\mathbf{r}_2$, and write the eigenstates of the Hamiltonian as $\Phi (\mathbf{R})\phi_{n}(\theta)$, where
\begin{eqnarray}
\phi_{n}(\theta)&=&\frac{1}{\sqrt{2\pi}}e^{in\theta}, \quad (n=0,\pm1,\pm2,\ldots),\nonumber\\
E_{n}&=&\frac{n^2}{ r^2}
\end{eqnarray}
Here, $r=|\mathbf{r}|$ and $\theta$ is the angle of rotation relative to the $x$-axis.

We introduce the random potential as follows:
\begin{eqnarray}
V(\mathbf{r}_1,\mathbf{r}_2)&=&\displaystyle\sum_{l,l'\in\mathbb{Z}}\lambda_{l,l'}(\delta(x_1-l)\delta (y_1-l')\nonumber\\
&&+\delta (x_2-l)\delta (y_2-l'))
\end{eqnarray}
where $\lambda_{l,l'}$ are random variables.

We  write $\mathbf{r}_1 =(x_1,y_1)=(x/2+r\cos\theta /2,y/2+r\sin\theta/2)$ and $\mathbf{r}_2=(x_2,y_2)=(x/2+r\cos (\theta+\pi)/2, y/2+r \sin (\theta +\pi)/2)$.

This leads to
\begin{eqnarray}
&&V(x,y,\theta)\nonumber\\
&&=\displaystyle\sum_{l,l'\in\mathbb{Z}}\lambda_{l,l'}\Bigr\{\delta (x+r\cos\theta -2l)\delta (y+r\sin\theta -2l')\nonumber\\
&&+\delta (x+r\cos (\theta +\pi)-2l)\delta (y+r\sin (\theta +\pi)-2l')\Bigr\}.
\end{eqnarray}

We expand the eigenstates of $H=H_0+V$ as
\begin{eqnarray}
\Psi &=&\displaystyle\sum_{n} u_{n}(\mathbf{R})\phi_{n}(\theta)=\displaystyle\sum_{n}u_{n}(x,y)\phi_{n}(\theta).
\end{eqnarray}
to write the time-independent Schr\"{o}dinger equation as follows:
\begin{eqnarray}\label{eq2}
&&-\frac{d^2 u_{n}(x,y)}{dx^2}-\frac{d^2 u_{n}(x,y)}{dy^2}+E_{n}u_{n}(x,y)\nonumber\\
&&+\displaystyle\sum_{m}V_{nm}(x,y)u_{m}(x,y)=Eu_{n}(x,y)
\end{eqnarray}
where
\begin{eqnarray}
V_{nm}(x,y)&=&\displaystyle\sum_{l,l'\in\mathbb{Z}}\lambda_{l,l'}(\phi_{n}^{*}(\theta_{xy})\phi_{m}(\theta_{xy})\nonumber\\
&&\qquad+\phi_{n}^{*}(\theta'_{xy})\phi_{m}(\theta'_{xy}))
\end{eqnarray}
with
\begin{eqnarray}
\theta_{xy}=\mbox{arctan}\left(\frac{2l'-y}{2l-x}\right),  \theta'_{xy}=\mbox{arctan}\left(\frac{2l'-y}{2l-x}\right)-\pi\nonumber\\
\end{eqnarray}
and $|2l'-y|\leq r$, $|2l-x|\leq r$.

By discretizing (\ref{eq2}) using finite-difference methods, we write the Hamiltonian 
\begin{eqnarray}\label{hamrotor}
H&=& J\sum_{x,y}(|x+1,y\rangle\langle x,y|+|x-1,y\rangle\langle x,y|\nonumber\\
&&|x,y+1\rangle\langle x,y|+|x,y-1\rangle\langle x,y|)\otimes I_{n}\nonumber\\
&&+(-4J+E_{n})I_{x}\otimes I_{y}\otimes I_{n}\nonumber\\
&&+\displaystyle\sum_{m,n, x,y}V_{nm}(x,y)|x,y,n\rangle\langle x,y, m|
\end{eqnarray}
acting on the state
\begin{eqnarray}
|\Psi\rangle =\displaystyle\sum_{x,y,n}u_{n}(x,y)|x,y,n\rangle
\end{eqnarray}
where we define
\begin{eqnarray}
|x,y,n\rangle =\displaystyle\sum_{\theta}\phi_{n}(\theta) |x,y,\theta\rangle =\hat{c}^{\dagger}_{x,y,n}|0\rangle
\end{eqnarray}
and $\hat{c}^{\dagger}_{x,y,n}$ is the creation operator of the rigid rotor with the translational position $\mathbf{R}=(x,y)$ and the rotational state $n$.

The Hamiltonian (\ref{hamrotor}) can then be written as
\begin{eqnarray}\label{hamsupp}
H&=&J\displaystyle\sum_{x,y,n}(\hat{c}^{\dagger}_{x+1,y,n}\hat{c}_{x,y,n}+\hat{c}_{x-1,y,n}^{\dagger}\hat{c}_{x,y,n}\nonumber\\
&&+\hat{c}^{\dagger}_{x,y+1,n}\hat{c}_{x,y,n}+\hat{c}^{\dagger}_{x,y-1,n}\hat{c}_{x,y,n})\nonumber\\
&&+\displaystyle\sum_{x,y,n}(-4J+E_{n})\hat{c}^{\dagger}_{x,y,n}\hat{c}_{x,y,n}\nonumber\\
&&+\displaystyle\sum_{m,n,x,y}V_{nm}(x,y)\hat{c}^{\dagger}_{x,y,n}\hat{c}_{x,y,m}.
\end{eqnarray}

 We can interpret this Hamiltonian as follows. As in the previous case with a harmonic oscillator, one can diagonalize (\ref{hamsub}) corresponding to a 2D disordered lattice with $U(|i-j|)=0$ if $|i-j|=r$ and $U(|i-j|)\rightarrow \infty$ otherwise, and select the eigenstates corresponding to the rigid rotor to study their localization. However, we can again reduce dimensionality by the following procedure.

If we introduce the translational coordinate $\mathbf{R}=(x,y)=(x_1+x_2, y_1+y_2)$ and the relative distance $\mathbf{r}= (\bar{x},\bar{y})=(x_1-x_2,y_1-y_2)$, we  write
\begin{eqnarray}\label{rotsub}
H&=&-8J'\displaystyle\sum_{x,y,\bar{x},\bar{y}}|x,y,\bar{x},\bar{y}\rangle\langle x,y,\bar{x},\bar{y} |\nonumber\\
&&+J'(\hat{\frak{R}}_{x}+\hat{\frak{R}}_{x}^{\dagger})(\hat{\frak{r}}_{\bar{x}}+\hat{\frak{r}}_{\bar{x}}^{\dagger})+J'(\hat{\frak{R}}_{y}+\hat{\frak{R}}_{y}^{\dagger})(\hat{\frak{r}}_{\bar{y}}+\hat{\frak{r}}_{\bar{y}}^{\dagger})\nonumber\\
&&+\displaystyle\sum_{x,y,\bar{x},\bar{y}}U(\sqrt{\bar{x}^2+\bar{y}^2})|x,y,\bar{x},\bar{y}\rangle\langle x,y,\bar{x},\bar{y}|\nonumber\\
&&+\displaystyle\sum_{l,l'}\lambda_{l,l'}(\delta (x+\bar{x}-2l)\delta (y+\bar{y}-2l')\nonumber\\
&&\qquad\qquad +\delta (x-\bar{x}-2l)\delta (y-\bar{y}-2l'))
\end{eqnarray}
where $\hat{\mathfrak{R}}_{x}=\displaystyle\sum_{x,y,\bar{x},\bar{y}}|x+1,y,\bar{x},\bar{y}\rangle\langle x,y,\bar{x},\bar{y}|$, $\hat{\mathfrak{r}}_x=\displaystyle\sum_{x,y,\bar{x},\bar{y}}|x,y,\bar{x}+1,\bar{y}\rangle\langle x,y,\bar{x},\bar{y}|$, $\hat{\mathfrak{R}}_{y}=\displaystyle\sum_{x,y,\bar{x},\bar{y}}|x,y+1,\bar{x},\bar{y}\rangle\langle x,y,\bar{x},\bar{y}|$, $\hat{\mathfrak{r}}_y=\displaystyle\sum_{x,y,\bar{x},\bar{y}}|x,y,\bar{x},\bar{y}+1\rangle\langle x,y,\bar{x},\bar{y}|$ and $U(\sqrt{\bar{x}^2+\bar{y}^2})=0$ if $\sqrt{\bar{x}^2+\bar{y}^2}=r$ and $U(\sqrt{\bar{x}^2+\bar{y}^2})\rightarrow \infty$ otherwise.

The first three terms (\ref{rotsub}) give the kinetic energy for a composite particle, $E_0=-8J'+4J'\cos aK_{x}\cos ak_{\bar{x}} +4J'\cos a K_{y}\cos ak_{\bar{y}}$ where $K_x$, $K_y$, $k_{\bar{x}}$, $k_{\bar{y}}$ are the wave vectors associated with the translational motion in $x, y$-directions, and the relative motion in $\bar{x},\bar{y}$-directions respectively. In the low energy limit, we can introduce the Taylor expansion to obtain  the parabolic dispersions with $K_x a, K_y a, k_{\bar{x}} a, k_{\bar{y}}a < \pi/2$:
\begin{eqnarray}
E_0\approx -2J' (a^2 K_x^2 +a^2 K_y^2 +a^2 k_{\bar{x}}^2 +a^2 k_{\bar{y}}^2).
\end{eqnarray}
  In this limit, the wavefunction for the rigid rotor on a lattice can be approximately obtained by the Hamiltonian in the continuum space (\ref{rotor0}) and the problem reduces to the one described by (\ref{hamsupp})  with $J'=J/2$ when we project the Hamiltonian onto the set of states of the rigid rotor $|x,y,n\rangle$.

For the internal rotational state, (\ref{hamsupp}) gives a good approximation when the wavelength $\bar{\lambda}=2\pi r/n > a$. In this article, we investigate the numerical results in the regime $K_x a, K_y a < \pi/2$. Furthermore the condition $2\pi r/n > a$ is satisfied for  the chosen $n$  with $r=2a$.

By diagonalizing (\ref{hamsupp}), we obtain the results shown in Fig. \ref{fig5} and Fig. \ref{fig6}. In the plot, the translational energy $\langle E_{R}\rangle =\langle H_{R}\rangle$ where $H_{R}$ is given by the first five terms of Eq. (\ref{hamsupp}). In Fig. \ref{fig5},  we observe that the distributions of $\xi$ and $\gamma$ exhibit similar behavior as depicted in Fig. 2 for the 1D oscillator in the main text. The coupling between the translational and rotational degrees of freedom thus weakens localization. (\ref{hamsupp})  effectively describes a (2+$\epsilon$)-dimensional system with $\epsilon$ accounting for internal states. As in the  example for the oscillator, the translational and rotational motions of the rotor dynamics are effectively uncoupled in the limit of large rotational excitation energy. In fact, Fig. \ref{fig6} shows that $\tilde{\xi}$ (i.e., the IPR without tracing out the internal degrees of freedom, $\tilde{\xi}=\displaystyle\sum_{\mathbf{R},\mathbf{n}}|\rho_{SE} (\mathbf{R},\mathbf{R};\mathbf{n},\mathbf{n})|^2$) takes a value close to $\xi$ as $1/r^2$ and purity increase, indicating that a $(2+\epsilon)$-dimensional problem is nearly reduced to a 2D problem for each internal state. On the other hand, when purity is small, $\tilde{\xi}$ is smaller than $\xi$ since the internal degrees of freedom become inseparable from the translational degrees of freedom, and the internal states act as   an additional dimension.

\end{document}